\documentstyle[11pt,paspconf,epsf]{article}

\begin{document}

\def\sles{\lower2pt\hbox{$\buildrel {\scriptstyle <} 
   \over {\scriptstyle\sim}$}}
\def\sgreat{\lower2pt\hbox{$\buildrel {\scriptstyle >} 
   \over {\scriptstyle\sim}$}}

\title{Advective Disks}

\author{R. Narayan}
\affil{Harvard-Smithsonian Center for Astrophysics,
    60 Garden Street, Cambridge, MA 02138, U.S.A.}

\begin{abstract}
Recent work on advection-dominated accretion flows (ADAFs) is
reviewed.  The article concentrates on an optically thin branch of
ADAFs which is present at mass accretion rates below a critical value
$\sim(10^{-2}-10^{-1})$ the Eddington rate.  Models based on this
branch have been quite successful at explaining a number of
low-luminosity X-ray binaries and galactic nuclei, and some brighter
systems.  Some progress has also been made toward understanding the
various spectral states of accreting black holes.  It is argued that
ADAFs may provide one of the best techniques for demonstrating the
reality of event horizons in black holes.
\end{abstract}


\keywords{black hole, accretion, advection, jets, X-ray binaries, active
galactic nuclei, event horizon}

\section{Introduction}

The physics of accretion is embodied in a set of basic equations which
describe the conservation of mass, momentum, and energy, coupled with
relations for viscosity, energy generation, and energy transfer.
Solutions of these equations are much valued since they can be used to
model systems we observe in astrophysics. Four solution branches have
been discovered and studied so far (Chen et al. 1995 discuss the
relationships among the branches):

\noindent
1. Thin accretion disk solution (Shakura \& Sunyaev 1973, Novikov \&
Thorne 1973): This classic solution is by far the best-known of the
four branches and has been used for modeling a large number of systems
(see Pringle 1981 and Frank, King \& Raine 1992 for reviews).

\noindent
2. Shapiro, Lightman \& Eardley (1976, hereafter SLE) solution: This
hot, optically thin branch makes use of a two-temperature plasma and
produces power-law spectra at X-ray and $\gamma$-ray energies,
consistent with observations of accreting black holes and neutron
stars.  However, the solution is violently unstable (Piran 1978).

\noindent
3. Optically thick advection-dominated solution (Abramowicz et
al. 1988): At sufficiently high mass accretion rates, a new branch of
solution is present where most of the radiation emitted by the
accreting gas is trapped in the flow and is advected into the central
star.  This regime of accretion was originally considered by Katz
(1977) and Begelman (1978), but a full analysis was first presented in
an important paper by Abramowicz et al. (1988).

\noindent
4. Optically thin advection-dominated solution (Narayan \& Yi 1994,
1995ab, Abramowicz et al. 1995): At sufficiently low mass accretion
rates, another advective solution branch occurs.  Here, the accreting
gas has such a low density that it is unable to cool efficiently, and
so the dissipated energy is advected into the central star as thermal
energy.  Advection of thermal energy is standard in models of
spherical accretion (Shvartsman 1971, Ipser \& Price 1977).  The idea
was mentioned in the context of rotating flows by Rees et al. (1982),
but this mode of accretion was established as an independent solution
branch only recently with the work of Narayan \& Yi (1994, 1995ab) and
Abramowicz et al. (1995).  (See Narayan 1996b for a review.)

The present article reviews advection-dominated accretion flows
(ADAFs), concentrating in particular on optically-thin ADAFs around
black holes.

\section{Dynamics of Accretion Flows}

\subsection {The Basic Equations}

Consider a steady axisymmetric accretion flow.  Mass conservation
gives
$$
\dot M=-4\pi RH\rho v=\,{\rm constant},
\qquad H=C_H(c_s/\Omega_K),\eqno (2.1)
$$
where $\dot M$ is the mass accretion rate, $R$ is the radius, $H$ is
the vertical scale height, $\rho$ is the mid-plane density, and $v$ is
the radial velocity.  $H$ is expressed in terms of the local
isothermal sound speed $(c_{s}^2 = p/\rho)$ and the Keplerian
frequency $ \Omega_{K} $, with a proportional constant $C_{H}$
($=(5/2)^{1/2}$ in the analysis of Narayan \& Yi 1995b, hereafter
NY95b).

In height-integrated form, the conservation laws of radial momentum,
angular momentum and energy give the following ordinary differential
equations:
$$
v{dv\over dR}=-\Omega_K^2R+\Omega^2R-{1\over\rho}{d\over dR}
(\rho c_s^2),\eqno (2.2)
$$
$$
{d\Omega\over dR}={v(\Omega R^2-j)\over R^2\nu},\eqno (2.3)
$$
$$
\rho vT{ds\over dR}=\rho\nu R^2\left({d\Omega\over dR}\right)^2
-q^-, \eqno (2.4)
$$
where $\nu$ is the kinematic coefficient of viscosity, $j\dot M$ is
the inward flux of angular momentum ($j$ is a constant independent of
$R$), $T$ is the temperature, $s$ is the entropy, and $q^{-}$ is the
radiative cooling rate per unit volume.  It is standard to write $\nu
=\alpha c_s^{2}/\Omega_{K}$, with $\alpha$ assumed to be a constant.

The calculation of $q^{-}$ requires a detailed analysis of radiation
emission mechanisms and radiative transfer. In this paper, we are
primarily interested in radiation processes in optically thin gases.

\subsection{Advection-Dominated Accretion}

Let us write the energy equation (2.4) compactly as follows:
$$
q^{\rm adv}=q^+-q^-, \eqno (2.5)
$$
where $q^{+}$ refers to the viscous heating of the gas and $q^{\rm
adv}$ is a term which describes advective transport of energy (usually
a form of cooling).

Equation (2.5) shows that three regimes of accretion are possible,
depending on the relative magnitudes of the three terms:

\begin{itemize}
\item $q^{+} \approx q^- \gg q^{\rm adv}$.  This corresponds to 
a cooling-dominated accretion flow, where the energy released via
viscous dissipation is mostly radiated locally.  The thin accretion
disk and the SLE solution belong to this regime.

\item $q^{\rm adv} \approx q^+ \gg q^-$.  This corresponds to the
two ADAF branches.  Most of the viscous energy is advected with the
gas.

\item $-q^{\rm adv} \approx q^{-} \gg q^{+}$: Here we have a cooling
flow (or Kelvin-Helmholtz contraction in the case of a star).  Energy
generation is negligible, but entropy is steadily converted to
radiation as the gas accretes.

\end{itemize}

\subsection{Dynamics of ADAFs --- Self-Similar Solutions}

Returning to the energy equation (2.4), let us write the right-hand
side as $f(R)q^{+}$, where $f(R) = (q^+-q^-)/q^{+}$.  The actual form
of $f$ depends on the details of the cooling, but in the specific case
of an ADAF we have $f(R) \to 1$ so that it is independent of $R$.
Narayan \& Yi (1994) showed that equations (2.1)--(2.4) then have an
exact analytical self-similar solution.  (A version of the solution
was discovered earlier by Spruit et al. 1987 in a different context.)
In the limit when $\alpha \ll 1$, the self-similar solution takes a
particularly simple form (see the original papers for the full
solution):
$$
v\approx -{3\alpha\over(5+2\epsilon)}\Omega_KR, 
\quad \Omega\approx\left({2\epsilon\over5+2\epsilon}\right)^{1/2}\Omega_K,
\quad c_s^2\approx{2\over(5+2\epsilon)}\Omega_K^2R^2, \eqno (2.6)
$$
where $\epsilon=(5/3-\gamma)/(\gamma-1)$, and $\gamma$ is the ratio
of specific heats.  This solution reveals several interesting features
of ADAFs.

Since $\gamma$ is likely to be in the range 4/3 to 5/3, $\epsilon$
lies in the range 0 to 1.  If $\alpha\ \sgreat\ 0.1$ as some arguments
seem to suggest (Narayan 1996a), then we see that the radial velocity
is comparable to the free-fall velocity, $v\sim v_{ff}=\Omega_{K}R$.
(The velocity is nevertheless subsonic.)  The gas thus accretes much
more rapidly than in a thin disk, where $v \sim \alpha
(H/R)^{2}v_{ff}$.  The angular velocity is distinctly sub-Keplerian,
and in fact tends to 0 as $\gamma \to 5/3$.  What holds the gas up
against gravity if it is sub-Keplerian?  The answer is the pressure
gradient, which can be quite large in ADAFs as seen from the fact that
$c_s^2/v_{ff}^2\sim0.3-0.4$.  The gas temperature is correspondingly
quite high; in fact, optically thin ADAFs have almost virial
temperatures.  This is not unexpected as all the energy released
during accretion remains in the gas as internal energy.

Since $c_{s} \sim v_{ff}$ the scale height is large: $H/R \sim 1$ (see
eq. 2.1).  Can the height-integrated equations (2.1)--(2.4) be trusted
under such extreme conditions?  Narayan \& Yi (1995a) considered a
general set of equations without height-integration and discovered
another self-similar ADAF solution.  Using this solution they showed
that the height-integrated equations are a surprisingly good
approximation to the exact problem even when $H \sim R$.

\subsection{Global Solutions}

To proceed beyond self-similarity, it is necessary to solve the full
equations (2.1)--(2.4) with proper boundary conditions.  This has been
done for the case of accretion on to a black hole by several groups
(Narayan, Kato \& Honma 1997a, Chen, Abramowicz \& Lasota 1997,
Nakamura et al. 1996).  These authors employ a pseudo-Newtonian
potential (Paczy\'nski \& Wiita 1980) to simulate a Schwarzschild
black hole, and obtain global transonic solutions for various choices
of $\alpha$ and $\gamma$.  They solve self-consistently for the sonic
radius $R _{s}$ (where the accreting gas goes supersonic) and the
parameter $j$.

When $\alpha\ \sgreat\ 0.01 $, the global solutions reveal some
unexpected features: (i) they remain sub-Keplerian at all radii, and
(ii) they have no pressure maximum outside the sonic radius.  Although
these results were already apparent in the work of Matsumoto, Kato \&
Fukue (1985) and Narayan \& Yi (1995a), they were nevertheless
considered a surprise.  Prior to the recent work, it had been widely
assumed that when an ADAF approaches close to a black hole it would
take the form of a thick torus with open funnels.  The new results,
however, suggest that ADAFs with moderately large $\alpha$ are not
toroidal at all; in fact, ADAFs may be more similar to slowly rotating
settling stars, as Fig. 1 shows.

\begin{figure}
\plotone{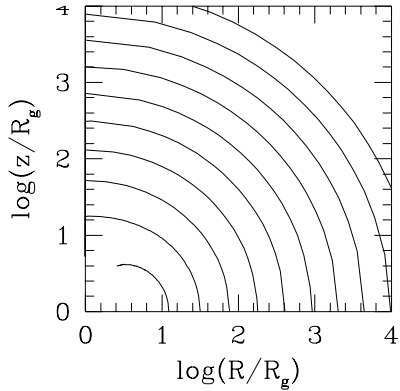}
\vspace{-3.4in}
\caption{Isodensity contours (two per decade of density) of a global
ADAF with $\alpha=0.3$ and $\gamma=1.5$ (based on Narayan et
al. 1997a, Appendix A).  The contours are truncated at the sonic
radius.  The axes are expressed in units of the Schwarzschild radius,
$R_g=2GM/c^2$.}
\end{figure}

Abramowicz et al. (1996) have extended the pseudo-Newtonian work to a
full general relativistic analysis and have presented global solutions
in Kerr geometry.  Save for some technical issues related to causality
in viscous interactions (see comments in Narayan et al. 1997a), it
appears that the dynamics of ADAFs within the height-integrated
approximation are now fully understood.

Much work remains to be done, however, on the two-dimensional
structure in the $Rz$ plane.  Figure 1 is only an approximate
representation, and real flows may well have complex circulation
patterns and jets which cannot be captured by the approximations used
so far.  

Narayan \& Yi (1994, 1995a) argued that ADAFs may be especially
susceptible to producing bipolar outflows, but no consistent solutions
have been found yet.  ADAFs are also convectively unstable, especially
along the axis (Narayan \& Yi 1994, 1995a)), as confirmed by
Igumenshchev et al. (1996).  The physical consequences of the
convection are not fully understood.

\subsection{Are There Shocks in ADAFs?}

The global solutions described in \S2.4 are free of shocks.  However,
Chakrabarti and his collaborators (cf. Chakrabarti \& Titarchuk 1995)
have claimed that shocks are generic to ADAFs.  Initially, Chakrabarti
(1990) considered viscous flows under isothermal conditions, but more
recently, with the increased interest in ADAFs, he has switched his
attention to adiabatic flows.  The clearest account of his results are
found in Chakrabarti \& Titarchuk (1995) where the authors claim: (i)
low $\alpha$ flows ($\alpha\ \sles\ 0.01$) have shocks and high
$\alpha$ ($\sgreat\ 0.01$) flows do not, and (ii) low $\alpha$ flows
have sub-Keplerian rotation at large radii while high $\alpha$ flows
are Keplerian at all radii except very close to the black hole.
Neither statement is confirmed in the work described in \S2.4.  In
particular, no shocks are seen in any of the global solutions, which
span a wide range of parameter values: $\alpha$ ranging from $10^{-3}$
to 0.3, and $\gamma$ from 4/3 to 5/3.

What is the source of the discrepancey?  Narayan et al. (1997a)
suggested that it may lie in different philosophies regarding the
angular momentum parameter $j$ in equation (2.3).  Chakrabarti simply
assigns a value to $j$ (in fact, different values in different papers)
whereas this parameter ought to be treated as an eigenvalue and
determined self-consistently through boundary conditions (see Popham
\& Narayan 1991 and Paczynski 1991).

In a viscous accretion flow the gas starts out with considerable
angular momentum at the outer edge, but very little of that angular
momentum is accreted by the central star since most of it is
transported outward by viscosity.  How much reaches the center is
determined by the properties of the star.  More specifically, there is
a radius near the inner edge where the flow satisfies a ``no-torque''
condition, and $j=\Omega R^{2}$ at this radius.  This well-known
argument leads in the case of a thin disk to the result $j =
\Omega_{K}(R_{in}) R_{in}^{2}$, where $R_{in}$ corresponds to the
inner edge of the Keplerian flow.  For more general global flows, it
is not possible to write an explicit formula for $j$ since neither the
sonic radius nor the run of $\Omega(R)$ is known ahead of time.
Instead, $j$ is determined self-consistently by requiring the solution
to satisfy boundary conditions on the inside appropriate to the
properties of the star.  Clearly, if one sets $j$ equal to an
arbitrary value the flow is likely to become self-inconsistent,
perhaps leading to a shock.  Indeed, self-inconsistency is likely
because the value of $j$ is known to be extremely insensitive to outer
boundary conditions (cf.  Narayan et al. 1997a), which means that a
randomly chosen value will not correspond to any reasonable flow.

At a deeper level, if $j$ is a global eigenvalue set by the star, how
does the flow on the outside know the value of $j$?  The answer is
that information is transported via viscous stresses (cf. Popham \&
Narayan 1992).  However, what happens when $\alpha$ is very small?  Is
it possible that when the viscous stress is weak information does not
make it to the outside, thus causing a shock?  No, because eq. (2.6)
shows that $v \propto \alpha$.  The smaller $\alpha$ gets, the slower
the radial velocity becomes, and the longer the gas spends within the
flow.  Therefore, regardless of the value of $\alpha$ information
always makes it back to the outside via viscosity, and the arguments
of the previous paragraph are always valid.  Some caveats are
necessary, however.  If $\alpha=0$ exactly, rather than $\alpha\to0$,
then there is no angular momentum transfer at all and shocks are
indeed generic.  Also, if the flow varies on a time scale shorter than
the viscous time $(R/v)$, shocks may be possible.  Finally, shocks do
occur when inflowing gas first meets a rotating flow, e.g. when the
accretion stream in a binary hits the disk.  We are not concerned with
such shocks.

\section{Optically Thin ADAFs}

Following Abramowicz et al. (1995), consider a simplified model where
electrons cool purely by free-free emission.  In such a gas, the
heating and cooling rates vary as $q^{+}\propto n_{i}$ and $q^-\propto
n_i n_e T_e^{1/2}$, where $n_i$ and $n_e$ are the ion and electron
number densities, and $T_e$ is the electron temperature.  Since $n_i
\sim n_e \propto \dot m$, it is clear that below a critical $\dot
m_{\rm crit}$, $q^{-}$ will be unable to match $q^{+}$ even if $T_e$
is equal to the local virial temperature $T_{vir}$.  The accretion
must then proceed as an ADAF.  (Here, $\dot m$ is the mass accretion
rate in Eddington units, $\dot m = \dot M/1.39\times10^{18}m ~{\rm
g\,s^{-1}}$, where $m$ is the black hole mass in solar mass units.)

In actual fact, once electrons become relativistic, they cool via
additional channels, notably synchrotron radiation and inverse Compton
scattering, which can be very efficient.  Can an ADAF survive in the
presence of these cooling processes?  Yes, because the viscous heating
primarily affects ions while the cooling is almost entirely by
electrons.  At low densities, the transfer of energy from ions to
electrons via Coulomb collisions is quite inefficient, creating a
bottleneck for efficient cooling.  In fact, under these conditions
the plasma is expected to become two-temperature with ion temperature
$T_{i}\gg T_e$ (SLE).  From NY95b, the rate of viscous heating of ions
scales as
$$
\left({dE_i\over dt}\right)_{visc}=9\alpha m^{-1}r^{-5/2}~{\rm
ergs\,s^{-1}}, \eqno (3.1)
$$ 
and the rate of energy transfer from ions to electrons
varies as
$$
\left({dE_e\over dt}\right)_{Coul}=6\alpha
m^{-1}\left({\dot m\over \alpha^2}\right)r^{-5/2}T_{10}^{-1}~{\rm
ergs\,s^{-1}}, \eqno (3.2) 
$$ 
where $r$ is the radius in Schwarzschild units and
$T_{10}=T_e/10^{10}$K (typically $T_{10}\sim0.1-1$).  From these
relations we see that the critical $\dot m$ below which a
two-temperature ADAF is possible is
$$ 
\dot m_{crit}\sim1.5T_{10}\alpha^2\sim(0.3-1)\alpha^2. \eqno (3.3) 
$$ 
This estimate is consistent with detailed calculations (NY95b,
Narayan 1996a).

Esin et al. (1996) have shown that even when there is efficient
coupling between ions and electrons (say by collective processes in
the plasma, cf. Begelman \& Chiueh 1988) ADAFs are still allowed, but
$\dot m_{crit}$ is much lower.

\subsection{Properties of Two-Temperature ADAFs}

The ion temperature in a two-temperature ADAF is very high,
$T_i\sim10^{12}{\rm K}/r$, and the electron temperature is also quite
high, $T_e\sim(10^9-10^{10})$K.  The optical depth to electron
scattering is usually low: $\tau_{es}\sim 10\alpha(\dot m/\dot
m_{crit})$ (NY95b).  These properties imply that such flows should
produce hard Comptonized spectra extending up to a few hundred keV.
ADAFs are therefore very attractive for modeling observations of
accreting black holes (\S\S4, 5).  Despite the high temperature, pair
processes are not important because the radiation density is low
(Bjornsson et al. 1996, Kusunose \& Mineshige 1996).

From eqs. (3.1), (3.2), we see that $q^-/q^+$ and the luminosity vary
as
$$
{q^-\over q^+}={(dE_e/dt)_{Coul}\over(dE_i/dt)_{visc}}
={\dot m\over \dot m_{crit}},
\quad {L\over L_{Edd}}\sim{q^-\over q^+}\dot m={\dot m^2\over\dot m_{crit}}.
\eqno (3.4)
$$
Low $\dot m$ ADAFs are thus intrinsically very inefficient, and the
efficiency drops rapidly with decreasing $\dot m$.  Not surprisingly,
the most successful applications of ADAF models have been to extremely
low-luminosity sources.

Where does all the energy in an ADAF go?  Ultimately, the advected
energy is accreted by the central star.  What happens then depends on
the nature of the star.  If we have a normal star with a surface, the
absorbed thermal energy will be re-radiated and the overall radiative
efficiency (when this re-radiation is also included) will be similar
to that of a standard thin accretion disk.  However, if the central
star is a black hole, then the advected thermal energy disappears
through the horizon and there is no re-radiation.  The overall
luminosity of the source is then significantly below what one expects
for the given $\dot m$.  This is a very specific signature of black
hole accretion.  Indeed, the firm demonstration that an accreting
source has an ADAF and that it is anomalously underluminous can be
construed as ``proof'' that the particular accreting star is a black
hole with an event horizon (Narayan, McClintock \& Yi 1996).  This is
one of the best astrophysical techniques available today to
demonstrate the reality of horizons.

\subsection{Stability}

Although the low $\dot m$ ADAF branch and the SLE branch are both hot
and optically thin, they differ in one important respect; the ADAF
solution is stable (Abramowicz et al. 1995, NY95b), whereas the SLE
solution is violently unstable (Piran 1978).

Consider vertically integrated quantities as a function of the ion
temperature $T_i$ in an accretion flow with a fixed $\dot m$.  The
heating rate per unit area, $Q^+=2Hq^+$, is directly proportional to
$\dot m$ and is independent of $T_i$.

At sufficiently low $T_i$, when the gas is optically thick (and
geometrically thin), the cooling rate $Q^-$ increases with increasing
$T_i$ because the temperature gradient goes up and the optical depth
goes down.  Beyond a certain critical temperature, however, the gas
goes optically thin and, with increasing $T_i$, the density decreases
and so does $Q^-$ (since the cooling varies as $\rho^2$).

Figure 2a shows a schematic plot of $Q^+$ and $Q^-$.  We see that two
solutions are in general possible which satisfy the condition
$Q^+=Q^-$ required for a cooling-dominated flow: (i) a cool optically
thick solution, and (ii) a hot optically thin solution.  The former
corresponds to the standard thin accretion disk.  This is a stable
solution (at least for fixed $\dot m$), in the sense that a slight
perturbation of $T_i$ causes the system to return to its equilibrium
state (indicated by the arrows in Fig. 2a).  The second solution
corresponds to the unstable SLE branch.

\begin{figure}
\plotone{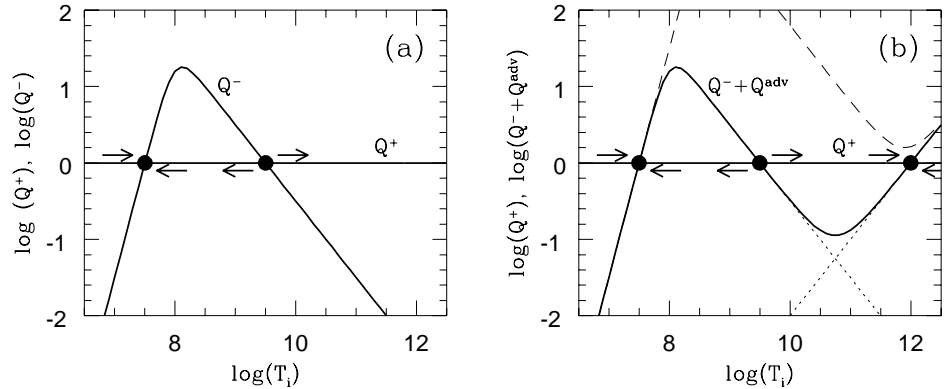}
\vspace{-3.3in}
\caption{(a) Shows $Q^+$ and $Q^-$ (normalized to $Q^+$) as functions
of $T_i$.  The two dots represent equilibrium cooling-dominated
solutions.  The arrows indicate the direction of evolution of a
non-equilibrium system; when $Q^+>Q^-$, the system heats up, and when
$Q^+<Q^-$ it cools.  The solution on the left (thin disk) is stable,
while the one on the right (SLE) is unstable.  (b) Includes advection.
A third solution is present, which is hotter than SLE, and stable.  The
dashed line corresponds to a system with $\dot m>\dot m_{crit}$.  Only
one solution survives.}
\end{figure}

The advective cooling term $Q^{\rm adv}$ is equal to the divergence of
the advective energy flux and is proportional to the local thermal
energy density of the gas.  Roughly we expect $Q^{\rm adv}\sim
Q^+(T_i/T_{vir})$ (see NY95b for more details), where $T_{vir}\sim
m_pc^2/kr$ is the virial temperature of the protons.  Thus, when
$T_i\sim T_{vir}$ it is possible to have an advection-dominated
solution satisfying $Q^{\rm adv}\sim Q^+$.

Figure 2b shows the situation when both cooling and advection are
present.  Now there are three solutions, of which two (the thin disk
and the ADAF) are stable, and one (SLE) is unstable.  At sufficiently
high $\dot m$ ($>\dot m_{crit}$) the SLE and ADAF solutions merge and
disappear and we are left with only the thin disk branch.  But for a
wide range of $\dot m<\dot m_{crit}$, both the thin disk and ADAF are
allowed.  Which of these two stable solutions does a system select?
NY95b discuss a number of possibilities; the current evidence suggests
that nature prefers the ADAF branch whenever $\dot m$ is low (\S4).

The above discussion refers to long-wavelength perturbations at
constant $\dot m$.  Kato, Abramowicz \& Chen (1996) have carried out a
more complete linear perturbation analysis and conclude that the low
$\dot m$ ADAF branch may have weakly unstable modes.  However, the
modes do not grow sufficiently rapidly to threaten the global
viability of the solution.  Indeed, Manmoto et al. (1996) use the
instabilities to explain X-ray shots in Cygnus X-1.

\section{Applications to Low Luminosity Systems}

ADAF solutions have been applied to several low luminosity black
holes, and model spectra have been fitted to observational data.
Detailed codes have been developed for this purpose which include
synchrotron and bremsstrahlung emission and Comptonization (see
Narayan, Barret \& McClintock 1997b for details).  Some earlier models
used the self-similar solution to estimate the run of density and
pressure (\S2.3), but the recent work employs global solutions
(\S2.4).

The models have been quite successful on two counts.  First, the
computed spectra are nearly always consistent with the available
observations, though the data are usually meager as the sources are
faint.  Second, some of the systems have independent estimates of
$\dot m$ and the sources appear to be extremely underluminous:
$L/L_{Edd} \sim(10^{-5}-10^{-3})\dot m$.  An ADAF can clearly explain
this.

The following subsections describe various applications of ADAF models
to low-luminosity systems.  To summarize, in every low-luminosity
system for which spectral data exist, the ADAF model provides a
plausible explanation of the observations.  Thus, for the moment at
least, it appears that the ADAF paradigm is well supported by
observations.  The evidence for event horizons may thus be considered
reasonably strong (\S3.1).

\subsection{Quiescent X-ray Binaries}

Narayan et al. (1996) showed that some puzzling observations of the
soft X-ray transient (SXT) source, A0620--00, in its quiescent state
could be explained with a model in which the accretion flow has two
zones: (i) an outer zone ($r>r_{tr}\sim10^3-10^4$) consisting of a
thin disk, and (ii) an inner zone ($r<r_{tr}$) consisting of an ADAF.
The model explains the available spectral data, and provides a
plausible explanation for the recurrence time between outbursts
(Lasota, Narayan \& Yi 1996).

Narayan et al. (1996) also modeled the quiescent SXT, V404 Cyg, and
predicted that it would have a hard X-ray spectrum with a photon index
$\sim2$.  This has now been confirmed with ASCA data (Narayan et al.
1997b).  Figure 3 shows a comparison between the model and
observations.  The fit is very good.  The model is quite robust and
fits the data for a wide range of black hole mass, binary inclination,
transition radius $r_{tr}$, viscosity $\alpha$, and magnetic field
strength.  The model also avoids several potential problems emphasized
by Wheeler (1996).

\begin{figure}
\plotone{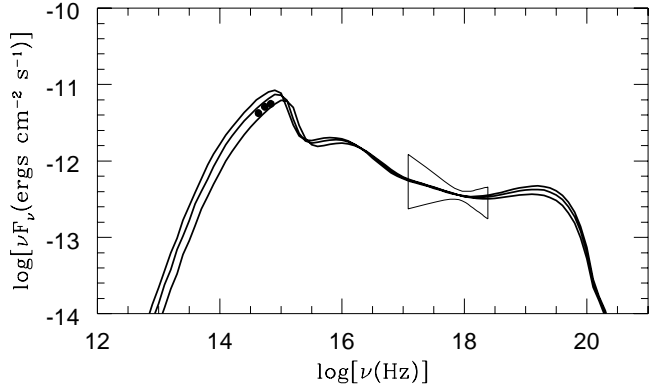}
\vspace{-3.3in}
\caption{The three dots show the optical flux of V404 Cyg in
quiescence, and the ``bow-tie'' is the X-ray error box ($2\sigma$)
from ASCA data.  The solid lines are the spectra of three ADAF-based
models with black hole masses (from left) of 16, 12, $8M_\odot$.
(From Narayan et al. 1997b)}
\end{figure}

V404 Cyg is the brightest known quiescent SXT and provides the best
test of the ADAF paradigm.  The evidence thus far is very encouraging.

\subsection{Quiescent Galactic Nuclei}

The first galactic nucleus to which an ADAF model was applied was
Sagittarius A$^*$ (Narayan, Yi \& Mahadevan 1995, see also Rees 1982),
the $\sim10^6M_\odot$ supermassive black hole at the center of our
Galaxy.  This source has long been known to be unusually dim for its
estimated mass accretion rate ($L/L_{Edd}\sim 10^{-5}\dot m$).  The
ADAF model provides a natural explanation of the low luminosity and
also gives a reasonable fit to various flux measurements and upper
limits of the source (ranging from radio to X-rays).

Following this work, Fabian \& Rees (1995) applied the same model to
other galaxies.  Fabian \& Canizares (1988) had earlier noted that the
nuclei of bright ellipticals are much too dim considering the mass
accretion rates estimated from cooling flows.  If the accretion is via
a thin accretion disk, the central black holes in these galaxies can
be no larger than $\sim(10^7-10^8)M_\odot$, which conflicts with the
much larger masses suggested by quasar evolution models.  This problem
can be reconciled if the black holes accrete through an ADAF (Fabian
\& Rees 1995, Mahadevan 1997).

\section{Applications to Luminous Systems}

\subsection{X-ray Binaries}

Narayan (1996b) constructed ADAF models with higher $\dot m$ than in
quiescent systems and found that systems in which $\dot m$ approaches
$\dot m_{crit}$ have very hard X-ray spectra with photon indices
$\sim1.5-2$.  This led to the plausible suggestion that the so-called
low state of black hole X-ray binaries may correspond to an ADAF flow
with $\dot m\ \sles\ \dot m_{crit}$.  Further, when $\dot m$ exceeds
$\dot m_{crit}$, the outer thin disk moves in to smaller radii and the
system switches to a pure thin accretion disk with a very soft
spectrum.  This corresponds to the high state.  Thus the transition
from low to high state corresponds to a switch from an ADAF to a thin
disk.  Chen \& Taam (1996) have come up with a somewhat different
proposal for the low-high transition.

Narayan \& McClintock (1997) carried out a more detailed test of these
ideas.  Using a sequence of models with varying $\dot m$ they tried to
explain the spectral evolution of Nova Muscae 1991 during outburst.
The results are fairly encouraging (Fig. 4).

From these studies, the following five states of an accreting black
hole can be identified, in order of increasing $\dot m$ (compare also
with Nowak 1995): (i) At very low $\dot m$, the ADAF extends over a
large range of $r$; if a thin disk is present at all, it is only at a
large radius; this is the quiescent state.  (ii) At higher $\dot m$,
but still with $\dot m<\dot m_{crit}$, the model geometry is
identical, but the luminosity is now much higher (recall eq. 3.4);
this is the low state.  (ii) For $\dot m\sim\dot m_{crit}$, the thin
disk moves in toward smaller radii, and we have a situation where the
ADAF extends over only a moderate range of $r$ and the thin disk is
energetically comparable to or even stronger than the ADAF; this is an
intermediate state.  (iv) At yet higher $\dot m$, the thin disk comes
all the way down to the last stable orbit and there is only a weak
corona; this is the high state.  (v) Finally, as $\dot m$ approaches
unity, the disk develops an energetically active corona (Haardt \&
Maraschi 1991), for reasons not yet understood; in this very high
state, the corona may be viewed as an ADAF which happens to coexist
with the thin disk.  (Alternatively, an ADAF is just a corona without
a disk.)

Ultra-relativistic jets appear to be associated with the very high
state (e.g. GRS 1915+105 and GRO J1655--40 discussed at this meeting).
The jets are most likely launched by the ADAF-like corona, which is
consistent with the idea that ADAFs are prone to produce jets (\S2.4).
We would then expect the intermediate state and low state also to have
jets, though these outflows may be less energetic.  Cyg X-1 and other
low state black holes with nonthermal radio emission may be examples
of this phenomenon.  (Quiescent systems too should have outflows, but
perhaps too weak to be observed.)

\begin{figure}
\plotone{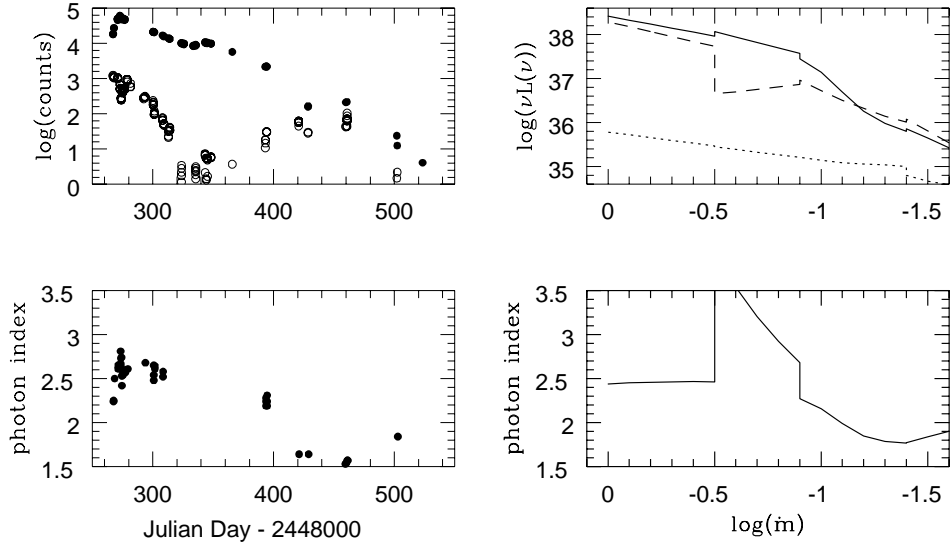}
\vspace{-2.4in}
\caption{The top left panel shows the X-ray counts of the SXT, Nova
Muscae 1991, during decline from outburst (Ebisawa et al. 1994); solid
dots: 1.2--4.6 keV, open dots: 9.3--37 keV.  The bottom left panel
shows the observed photon spectral indices; the spectral index could
not be measured between Day 320 and Day 390.  The top right panel
shows model fluxes (solid line: 1--6 keV, dashed: 6--40 keV, dotted:
2--10 eV), and the bottom right panel shows model spectral indices.
The models reproduce the qualitative features seen in the data.  (From
Narayan \& McClintock 1997) }
\end{figure}

\subsection{Galactic Nuclei}

Lasota et al. (1996) made an ADAF model of NGC 4258 and showed that it
explains the observed X-ray spectrum.  Herrnstein et al.  (this
volume) show that the same model is also consistent with VLBI radio
observations.  Since the model gives a reasonably large $\dot
m\sim0.01$, NGC 4258 corresponds to a state that is in between the
quiescent and low states of X-ray binaries.  NGC 4258 is a LINER
galaxy and Lasota et al. suggested that perhaps all LINERs have ADAFs.
In terms of $\dot m$, LINERS probably range all the way from the
quiescent state up to the intermediate state.

Carrying through the analogy with X-ray binaries, one expects that
active galactic nuclei too should have multiple states.  We have
already mentioned the quiescent, low and intermediate states above.
Seyfert galaxies probably span the range from the intermediate state
to the low end of the high state, while bright quasars may correspond
to the high and very high states.  Again, in analogy with X-ray
binaries, we would expect the most energetic jets to occur in quasars
in the very high state (these would correspond to FR II systems),
while we would have lower power jets in systems corresponding to the
low and intermediate states (FR I).  In this picture, radio quiet
quasars would correspond to the high state, where there is neither a
separate ADAF nor an active corona; in terms of $\dot m$, radio quiet
quasars should lie in between FR II and FR I systems.

Yi (1996) has argued that the break in the quasar luminosity function
at redshift $z\sim2$ might be the result of quasars switching from
thin disk accretion (at $z>2$) to ADAF accretion ($z<2$).  Such an
evolution is expected just from the increase of black hole mass with
time, aided perhaps by a decrease in the mass accretion rate.  The
model naturally explains the lack of bright quasars at the present
epoch.

I thank Marek Abramowicz, Jean-Pierre Lasota, Jeff McClintock and Insu
Yi for comments.  This work was supported in part by NASA grant NAG
5-2837.

\begin{question}{Dr.\ Michael Nowak}
Two questions: In the low state of Galactic black hole candidates,
what is the dominant source of soft photons for Comptonization?  Do
you include the irradiation of the cool disk from the Compton cloud?
Second, you predict that the low-to-high state transition is the
optically thick disk working its way inward, whereas other theories
may predict the opposite (an optically thick disk forming in the
center and working its way outward).  Can this be distinguished with
the recent Cyg X-1 high state observations?  Can you think of other
tests?
\end{question}
\begin{answer}{Dr.\ Narayan}
(1) In the low state as well as the quiescent state, the dominant
source of soft photons is synchrotron radiation from within the ADAF.
We do include irradiation of the outer disk by radiation from the
inner ADAF, as well as irradiation of the ADAF by the cool disk.
These interactions are important in the intermediate state, but not in
the quiescent and low states.  (2) We have not tried to model the
recent Cyg X-1 high state data, but there are several tests of our
ideas that are potentially possible with SXTs.  We predict that during
the rise phase of an SXT outburst the inner edge of the thin disk
moves in, and that late in the decline, once the system switches from
the high state to the intermediate state, the disk moves out.  The
variation of the inner edge should be seen in multi-color blackbody
fits of the soft component of the spectrum.  If an X-ray iron line is
seen, this should show the variation of the Keplerian velocity at the
inner edge.  Also, we have clear predictions for the variation of the
X-ray reflection component.
\end{answer}

\begin{question}{Dr.\ Andrew King}
Two questions: (1) What causes the SXT outburst in your model?  (2) In
the case when the accreting object is a neutron star, am I right in
thinking that the advective flow will crash into the surface and give
back all the advected energy?  So these systems cannot be ``secret
guzzlers''?
\end{question}
\begin{answer}{Dr.\ Narayan}
(1) I believe the outburst is caused by a thermal limit cycle in the
outer thin disk, exactly as in current models (e.g.  Mineshige \&
Wheeler, ApJ, 343, 241, 1989).  However, the fact that the disk is
truncated at a large transition radius $r_{tr}$ will doubtless cause
important changes (cf. Lasota, Narayan \& Yi 1996) and needs to be
modeled in detail.  (2) Yes, there ought to be a clear difference
between black hole SXTs and neutron star SXTs, especially in
quiescence, since the former are ``secret guzzlers'' (cf. Abramowicz
\& Lasota, Comments Ap, 18, 141, 1995, who introduced the term)
according to our model while the latter cannot be secret guzzlers (see
\S3.1).  There is some evidence that, in quiescence, Cen X-4 (a NS
SXT) is brighter than A0620--00 (a BH SXT), which is consistent with
this expectation.  However, detailed ADAF models have not yet been
developed for NS SXTs and so it is not clear if the observed spectra
can be fitted.
\end{answer}

\begin{question}{Dr.\ Mario Livio}
Your models do not produce a secondary maximum in the optical-UV,
while these are observed in many systems.
\end{question}
\begin{answer}{Dr.\ Narayan}
Our models do not show any effect in the optical-UV band (dotted line
in Fig. 4) when the soft and hard X-rays go through their ``glitch.''
The X-rays are from the ADAF/corona, and the optical-UV are from the
cool disk.  Perhaps we could make our model produce a secondary
maximum in the optical light curve if we change the way we model the
irradiation of the disk.
\end{answer}

\begin{question}{Dr.\ Mitch Begelman}
In your model for the different spectral states of black hole
candidates, a transition from low state to high state should be
accompanied by a large increase in $L_{bol}$, since $\dot M$ is
increasing and the advection-dominated zone is shrinking.  But in data
shown by Zhang this morning, $L$(1--200 keV) was constant during the
transition of Cyg X-1 from low state to high state.  Can this be
accommodated within your scheme?
\end{question}
\begin{answer}{Dr.\ Narayan}
I believe the Cygnus X-1 ``high state'' is not a true high state but
more like what I have called the intermediate state.  We would predict
an increase in $L_{bol}$ from the low state to this state, exactly as
you say.  However, if the thin disk remains at a moderately large
radius, then much of its radiation may lie outside the 1-200 keV band,
and the observed $L$ in the X-ray band may well remain constant even
while $L_{bol}$ is going up.  In our model of SXT outbursts, for
instance, during the rise phase (not discussed in the article) we
actually see the X-ray flux {\it fall} briefly during a period when
the bolometric luminosity is rising.  So, Zhang's observations could
be accommodated in principle, but we would need to model his data in
detail before we can tell for sure.
\end{answer}

\end{document}